\documentclass{nature_sub}
\usepackage[utf8]{inputenc}
\usepackage{xcolor,ulem}
\usepackage{graphicx}
\usepackage{amssymb}
\usepackage{amsmath}
\usepackage{bm}
\usepackage{color}
\usepackage{hyperref}
\usepackage{verbatim}

\newcommand{\nat}{Nature}

\newcommand{\mnras}{MNRAS}
\newcommand{\apj}{Astrophys. J.}

\newcommand{\apjl}{Astrophys. J. Lett.}

\newcommand{\prd}{Phys. Rev. D}
\newcommand{\prl}{Phys. Rev. Lett.}

\newcommand{\sint}{{\rm int}}

\title{AGN as Potential Factories for Eccentric Black Hole Mergers}

\author{J. Samsing$^{1}$\footnote{Corresponding author}, I. Bartos$^{2}$, D. J. D'Orazio$^{1}$, Z. Haiman$^{3}$, B. Kocsis$^{4}$, N. W. C. Leigh$^{5,6}$, B. Liu$^{1}$, M. E. Pessah$^{1}$, H. Tagawa$^{7}$.}

\begin{document}


\maketitle

\begin{affiliations}
\begin{small}
 \item Niels Bohr International Academy, The Niels Bohr Institute, Blegdamsvej 17, DK-2100, Copenhagen \O, Denmark.
 \item Department of Physics, University of Florida, PO Box 118440, Gainesville, FL 32611, USA.
 \item Department of Astronomy, Columbia University, 550 W. 120th St., New York, NY, 10027, USA.
 \item Rudolf Peierls Centre for Theoretical Physics, Clarendon Laboratory, Parks Road, Oxford OX1 3PU, UK.
 \item Departamento de Astronom\'ia, Facultad Ciencias F\'isicas y Matem\'aticas, Universidad de Concepci\'on, Av. Esteban Iturra s/n Barrio Universitario, Casilla 160-C, Concepci\'on, Chile.
 \item Department of Astrophysics, AMNH, Central Park West at 79th Street, New York, NY 10024, USA.
 \item Astronomical Institute, Tohoku University, Sendai, Miyagi 980-8578, Japan.
\end{small}
\end{affiliations}

\begin{abstract}

There is some weak evidence that the black hole merger named GW190521 had a non-zero
eccentricity\cite{2020arXiv200905461G, 2020ApJ...903L...5R}. In addition, the component black holes' masses exceeded the limit predicted by stellar evolution\cite{GW190521properties}. The large masses can be explained by successive mergers\cite{2019PhRvL.123r1101Y,2020ApJ...898...25T}, which may be efficient in gas disks surrounding active galactic nuclei (AGN), but it is difficult to maintain an eccentric orbit all the way to the merger, as basic physics would argue for circularization\cite{Peters:1964bc}. Here we show that AGN-disk environments can lead to an excess of eccentric mergers, if the interactions between single and binary black holes are frequent\cite{2020ApJ...898...25T}, and occur with mutual inclinations of less than a few degrees. We further illustrate that this eccentric population has a different distribution of the inclination between the spin vectors of the black holes and their
orbital angular momentum at merger\cite{2020PhRvL.125j1102A}, referred to as the {\it spin-orbit tilt}, compared to the remaining circular mergers.

\end{abstract}

\begin{figure}
\centering
\includegraphics[width=\textwidth]{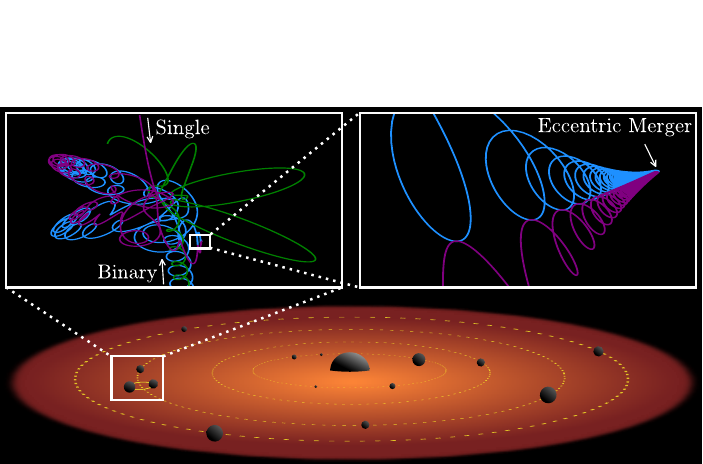}
\caption{{\bf Illustration of an eccentric LIGO-Virgo source forming in an AGN-disk.}
{\it Bottom:} AGN-disk (not to scale) with its central super-massive black hole, and a population of smaller orbiting black holes. These smaller black holes occasionally pair-up to form binary black holes, which often undergo scatterings with the single black hole population.
{\it Top:} Outcome of a $[50M_{\odot}, 80M_{\odot}]$ binary black hole interacting with an incoming $[70M_{\odot}]$ black hole that results in a
$[80M_{\odot}, 70M_{\odot}]$ binary black hole merger during the interaction with an eccentricity $\sim 0.5$ in LIGO-Virgo.}
\label{fig:AGNill}
\end{figure}

Black holes that eventually merge in AGN-disks can be brought into the disk through gas-capture from the surrounding nuclear star cluster\cite{2017ApJ...835..165B} or can be produced through in-situ star formation\cite{2003astro.ph..7084L,2017MNRAS.464..946S,2020arXiv200903936C}. 
Once a black hole is in the disk, it will undergo radial migration\cite{2003astro.ph..7084L}, and can as a result pair up with another black hole to form a binary\cite{2012MNRAS.425..460M, 2018ApJ...866...66M}. Recent studies show that interactions between such migrating binary black holes and other single black holes in the AGN-disk, referred to as {\it binary-single interactions},
likely provide the main pathway for bringing binaries to merger\cite{2018MNRAS.474.5672L, 2019ApJ...878...85S, 2020ApJ...898...25T,2020ApJ...899...26T,2006ApJ...648..411K} (see Fig.~\ref{fig:AGNill}).
Despite progress on characterizing such interactions\cite{2018MNRAS.474.5672L}, the inclusion of gravitational wave emission {\it during} the interactions,
which has been shown to be essential in resolving mergers with non-zero eccentricity that form in stellar-clusters\cite{2018PhRvD..98l3005R, 2018PhRvD..97j3014S}, remains unexplored. 
Observationally, GW190521 is among the first gravitational wave sources with indications of an AGN-disk origin\cite{2020PhRvL.124y1102G,GW190521properties}. It is sensible to inquire whether its apparent non-zero eccentricity\cite{2020arXiv200904771R, 2020arXiv200905461G}, as well as its observed $\sim 90^\circ$ spin-orbit tilt\cite{2020PhRvL.125j1102A}, could arise naturally as a distinct signature, characteristic of dynamically induced AGN-disk mergers.

\begin{figure}
\centering
\includegraphics[width=0.95\columnwidth]{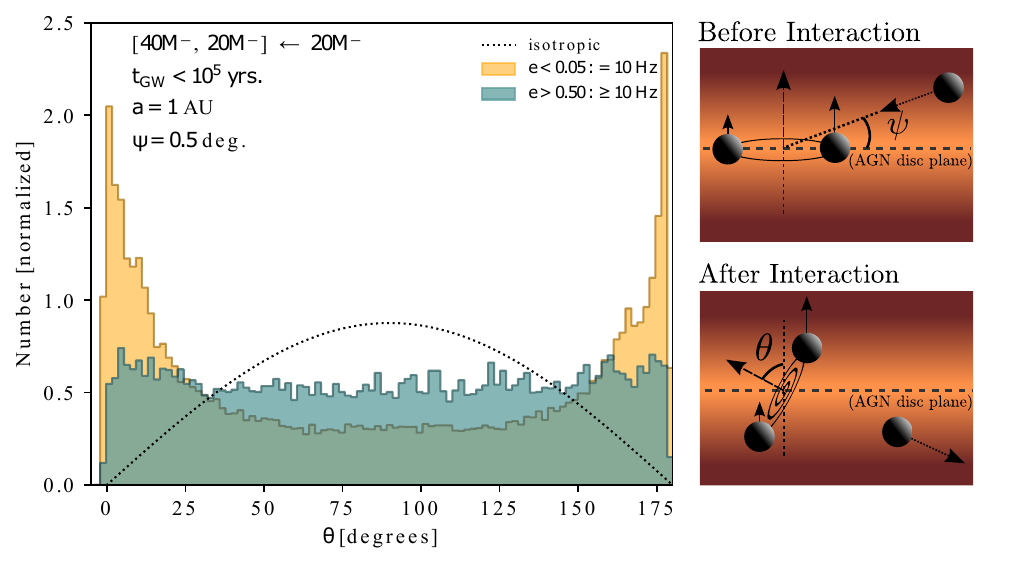}
\caption{{\bf Distribution of Orbital Plane Orientations.} 
Results from 2.5-PN scatterings between an incoming $20M_{\odot}$ black hole and a $[40M_{\odot}, 20M_{\odot}]$ binary black hole
with $a = 1\ {\rm AU}$. The initial orbital inclination (depicted in top right panel) is set to $\psi = 0.5^\circ$.
The histograms, each normalized, show the inclination angle between the orbital angular momentum vector of the
final merging binary black hole and the initial one, $\theta$ (depicted in lower right panel), where {\it yellow} and {\it blue} refer to the populations of merging binaries
with eccentricity $e<0.05$ at $10$~Hz and $>0.5$ at $\geq 10$~Hz, respectively. The {\it dotted line} shows for comparison an isotropic distribution expected in the 3D cluster-like case.}
\label{fig:angle_dist}
\end{figure}

With this motivation, we explore how binary black holes merge through binary-single interactions in AGN-disk environments when gravitational wave emission is 
included in the dynamics via the 2.5-post-Newtonian (2.5-PN) term\cite{Blanchet:2006kp} (See Methods).
To approach this complex astrophysical situation in a systematic way, we focus here on quantifying the unique signatures that might be associated with the approximate 2-dimensional (2D) disk-like environment of the AGN-disk
compared to the usual 3-dimensional (3D) interactions found in stellar clusters\cite{2018PhRvD..97j3014S}. For this, we perform controlled experiments of initially circular black hole binaries interacting with singles incoming on an orbital plane that is inclined relative to the binary orbital plane by an angle $\psi$, where $\psi = 0$ corresponds to a co-planar interaction (Fig.~\ref{fig:angle_dist}).
For a given scattering, we study the characteristics of the merging black holes, which we divide into the following two distinct categories;
{\it $3$-body merger}: two of the three interacting black holes merge while they are all bound and interacting\cite{2018PhRvD..97j3014S} (see Fig. \ref{fig:AGNill}).
{\it $2$-body merger}: the binary black hole survives the three-body interaction, but merges before undergoing its next interaction\cite{2018MNRAS.tmp.2223S}.  

\begin{figure}
\centering
\includegraphics[width=0.90\columnwidth]{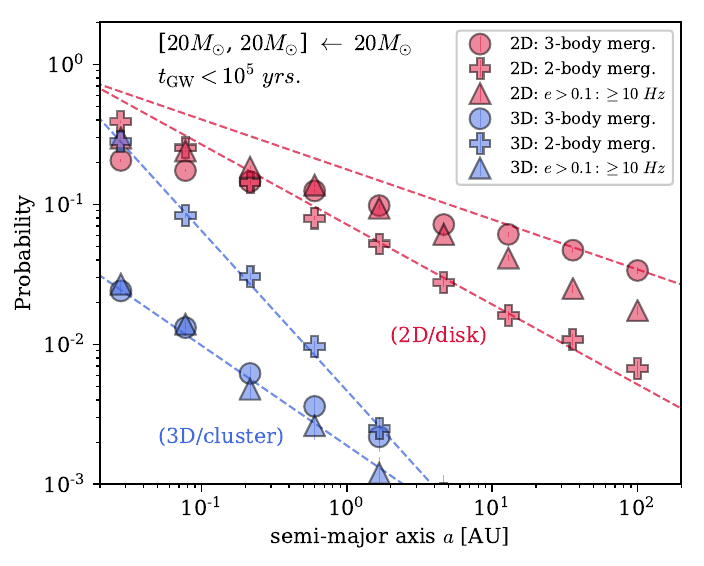}
\caption{{\bf Merger Probability.} Probability of undergoing a $2$-body merger (Plus-signs) or a $3$-body merger (Circles) with inspiral time $t_{GW} < t_{\sint} = 10^{5}\ {\rm yr}$, derived from 2.5-PN binary-single interactions
between black holes with equal mass $20M_{\odot}$, as a function of initial semi-major axis, $a$. The Triangles show the probability of undergoing a merger with $e > 0.1$ at $\geq 10$~Hz. Red, numerical results from
2D co-planar interactions. Blue, numerical results from 3D isotropic interactions. Dashed lines, analytical approximations (see Methods).}
\label{fig:P23_a}
\end{figure}

Fig. \ref{fig:P23_a} shows the probability for merger as a function of black hole binary semi-major axis, $a$. As seen in this figure, restricting the interactions to be co-planar leads to a significant enhancement of mergers; for example, for a binary with a semi-major axis $a\sim 1$~AU the fraction of 3-body mergers is $\sim 100$ times larger in the 2D disk-case compared to the 3D cluster-case. As outlined in the Methods, this enhancement is due to the difference in eccentricity distributions of the dynamically assembled binaries, $P(e)$, that follows $\approx e/\sqrt{1-e^2}$ in the 2D case compared to $\approx 2e$ in
the 3D case\cite{1976MNRAS.177..583M, 2006tbp..book.....V, 2019Natur.576..406S}.
Our analytic approximations (see Methods) for both the 2-body- and 3-body merger probabilities, $p_2$ and $p_3$, respectively, are also included in Fig. \ref{fig:P23_a}. Assuming the equal-mass limit, the corresponding ratio of probabilities between the 2D and 3D cases is,
\begin{align}
\frac{p_2^{\rm (2D)}}{p_2^{\rm (3D)}}  & \approx 10^1 \times   \left[\frac{m}{20M_{\odot}}\right]^{-6/14}\left[\frac{a}{\rm 1\ {\rm AU}}\right]^{8/14}\left[\frac{t_{\sint}}{10^{5}\ \mathrm{yr}.}\right]^{-1/7}, \\
\frac{p_3^{\rm (2D)}}{p_3^{\rm (3D)}}  & \approx 10^2 \times \left[\frac{m}{20M_{\odot}}\right]^{-5/14}\left[\frac{a}{\rm 1\ {\rm AU}}\right]^{5/14},
\end{align}
where $m$ is the black hole mass, and $t_{\sint}$ is the time in-between interactions scaled to a value characteristic for AGN-disk models\cite{2020ApJ...898...25T}.
These results show that the effects from breaking the scattering isotropy become increasingly important 
at larger $a$ and smaller $m$. Note further the weak dependence on $t_{\sint}$.

An interesting observational consequence of the enhancement of the merger probabilities is the increase in the number of black hole mergers with a measurable eccentricity, $e$, at gravitational wave frequency $f_{GW}$.
This is illustrated in Fig. \ref{fig:P23_a}, which includes the probability that a given scattering results in a merger with a
measurable $e$ in LIGO-Virgo ($e> 0.1$ at $f_{GW} \geq 10\ \mathrm{Hz}$). As seen in the figure, the eccentric population follows closely the 3-body merger population, but the scaling with $m$ and $a$ is slightly 
different. As a consequence, the probability of forming a merger with $e> 0.1$ at $f_{GW} \geq 10\ \mathrm{Hz}$ in the 2D case is enhanced relative to the 3D case by the factor (see Methods)
\begin{equation}
\frac{p_{0.1,10 \mathrm{Hz}}^{(\rm 2D)}}{p_{0.1,10 \mathrm{Hz}}^{(\rm 3D)}} \approx 10^2 \times \left[\frac{m}{20M_{\odot}}\right]^{-1/6}\left[\frac{a}{\rm 1\ {\rm AU}}\right]^{1/2}.
\end{equation}

\begin{figure}
\centering
\includegraphics[width=0.75\columnwidth]{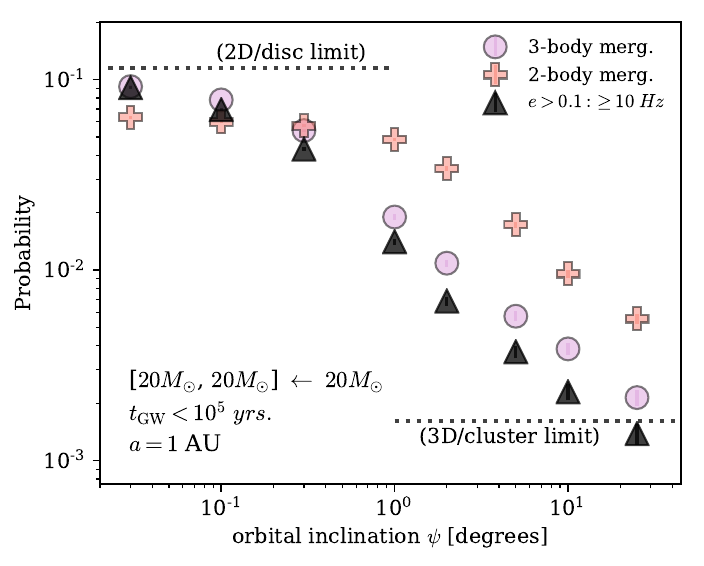}
\caption{{\bf Dependence on Orbital Inclination.} 
Probability of undergoing a $2$-body merger (Plus-signs) or a $3$-body merger (Circles) with inspiral time $t_{GW} < t_{\sint} = 10^{5}\ {\rm yr}$, derived from 2.5-PN interactions
between a binary with $a = 1$ AU and an incoming single, for varying relative orbital inclination angle $\psi$. The Triangles show the probability of undergoing a merger with $e > 0.1$ at $\geq 10$~Hz. 
All three objects are black holes with equal mass $m = 20M_{\odot}$.
The dotted lines illustrate the 2D (disk) and 3D (cluster) limits for the eccentric population ($e > 0.1$ at $\geq 10$~Hz).}
\label{fig:ecc_thetaz_i}
\end{figure}

Fig. \ref{fig:ecc_thetaz_i} extends this analysis to the less-idealized case
by plotting the merger probabilities as a function of $\psi$ for fixed $a = 1$~AU. As seen in the figure, $\psi$ must here be $\lesssim 1^{\circ}$ for the boost in eccentric LIGO-Virgo mergers to be notable.
The true distribution of $\psi$ is observationally unconstrained, and it is unclear whether binary-single scatterings are within a few degrees of coplanar, as is necessary in our model. If the characteristic inclination angle is comparable to the AGN-disk thickness $h$ divided by the radius $r_H \approx R(m/M)^{1/3}$ of the Hill sphere, where $R$ is the distance to the central super massive black hole with mass $M$,
then $\psi \approx 1^\circ$ for thin disk models\cite{2005ApJ...630..167T, 2020ApJ...898...25T} ($h/R \sim 10^{-3}$, $M \sim 10^6 M_{\odot}$, $m \sim 50 M_{\odot}$). If instead the characteristic inclination angle is given by $h$ divided by the semimajor axis of the binary (as might happen if turbulent eddies and/or two-body scattering are important), then $\psi\approx 1$~radian or more and interactions co-planar enough for our mechanism to operate are rare. It is uncertain whether gas drag will typically lead to efficient co-planar alignment, but if it does then $\psi<{\rm few}$ degrees would be less rare.

Finally, our 2.5-PN scatterings reveal also a new interesting relation between black hole merger eccentricity and spin-orbit tilt caused by the anisotropic environment provided by the AGN-disk.
The spin-orbit tilt is a powerful statistical measure of the underlying formation channel, e.g. sources dynamically assembled in stellar clusters are expected to have an isotropic
distribution in contrast to mergers originating from isolated stellar binary evolution\cite{2016ApJ...832L...2R}.
Fig.~\ref{fig:angle_dist} explores this for AGN-disk mediated mergers, by plotting distributions of the angle between the orbital angular momentum vector of the final merging binary black hole and that of the initial black hole binary, denoted $\theta$, with Fig.~\ref{fig:fGW_dist} (Methods) showing the corresponding distributions of gravitational wave frequency and eccentricity.
We split the distributions in Fig.~\ref{fig:angle_dist} into two categories based on binary eccentricity at the time of detectability, $e<0.05$ at $ = 10$~Hz (yellow) and $e>0.5$ at $\geq 10$~Hz (blue).
As seen in the figure, the distribution of $\theta$ with notable eccentricity in LIGO-Virgo ($e>0.5$-population) is much broader and flatter than for the
remaining circular ($e<0.05$-population).
Part of the reason is that eccentric mergers have a relatively small angular momentum vector, 
which can therefore be easily re-oriented without being significantly constrained by the (near co-planar) initial conditions, in contrast to the less eccentric.
The angle $\theta$ is not directly observable; however, the gas that brings the interacting black holes close to the plane of the AGN-disk prior to interaction\cite{2020ApJ...898...25T} will also spin them up through accretion, which will drive their spin vectors towards being perpendicular to the AGN-disk. As a result, $\theta$ is approximately equal to the observable spin-orbit tilt. This opens up the novel possibility of using eccentricity together with spin-orbit tilt to probe the origin of AGN-disk mediated mergers, and provides further a possible explanation for the non-zero eccentricity and noticeable spin-orbit tilt of GW190521 if it formed in an AGN-disk.

\begin{addendum}
\item[Acknowledgments] The authors are grateful to Nirban Bose, Kelley Holley-Bockelmann, Archana Pai, Mike Zevin and M. Safarzadeh for their useful suggestions.
J.S. is supported by the
European Union's Horizon 2020 research and innovation programme under the Marie Sklodowska-Curie grant agreement No. 844629.
D.J.D received funding from the European Union's Horizon 2020 research and innovation programme under the Marie Sklodowska-Curie grant agreement No. 101029157 and through Villum Fonden grant No. 29466.
I.B. acknowledges support from the Alfred P. Sloan Foundation.
H.T. is financially supported by the Grants-in-Aid for Basic Research by the Ministry of Education, Science and Culture of Japan (HT:17H01102, 17H06360) and Japan Society for the Promotion of Science (JSPS) KAKENHI 
Grant Number JP21J00794 (HT). 
N.W.C.L. gratefully acknowledges the generous support of a Fondecyt Iniciaci\'on grant 11180005, and the financial support from Millenium Nucleus NCN19-058 (TITANs) and the BASAL Centro de Excelencia en Astrofisica y Tecnologias Afines (CATA) grant PFB-06/2007.
Z.H. acknowledges support from NASA grant NNX15AB19G and NSF grants AST-1715661 and AST-2006176. This project has received funding from the European Research Council (ERC) under the European Union's Horizon 2020 research and innovation programme ERC-2014-STG under grant agreement No 638435 (GalNUC) to BK.
\item[Author contributions]
J.S. led the work, carried out the simulations and calculations, and wrote the initial manuscript together with I.B. and D.J.D.
The remaining authors, Z.H, B.K., N.W.C.L, B.L, M.P. and H.T. all contributed equally to the intellectual development of the ideas
and the preparation of the final manuscript.
\item[Competing Interests] The authors declare no competing financial interests.
\item[Correspondence] Correspondence and requests for materials should be addressed to J. Samsing  (email:\\jsamsing@gmail.com)
\item[Code availability] Requests for codes should be addressed to J. Samsing  (email:\\jsamsing@gmail.com)
Reprints and permissions information is available at www.nature.com/reprints
\end{addendum}

\newpage

\clearpage

\section*{Methods}

\section{Analytical Estimations and Theoretical Modeling}

The co-planar 2-dimensional (2D) binary-single scattering setup, for which $\psi = 0$, not only represents the limiting case of our AGN-disk scattering model,
but leads also to an upper limit on the number of mergers compared to the isotropic 3-dimensional (3D) cluster case. Although idealized, it is therefore useful to consider this
model to explore the expected maximum enhancement of the merger probability, when going from the cluster-like 3D case to the disk-like 2D case.
In the following analytical treatment we assume equal-mass binary components and omit the possible effects from gas-drag in the dynamics\cite{1999ApJ...513..252O, 2007ApJ...665..432K}.

We start by calculating the probability $P(t_{GW} < \tau)$ that a dynamically assembled binary black hole has a gravitational wave inspiral time $t_{GW}$ less than some timescale $\tau$.
For an eccentric binary black hole, $t_{GW}$ is given approximately by\cite{Peters:1964bc}
\begin{equation}
t_{GW} \approx  \frac{5c^5}{512G^3} \frac{a^4}{m^3} (1-e^2)^{7/2} \approx t_c \times (1-e^2)^{7/2},
\end{equation}
where $e$, $a$, and $m$ are the initial orbital eccentricity, semi-major axis, and individual black hole mass, respectively, and $t_c$ denotes the circular inspiral time for which $e=0$.
The eccentricities of binary black holes dynamically assembled through 2D co-planar interactions are distributed
as\cite{1976MNRAS.177..583M, 2006tbp..book.....V, 2019Natur.576..406S}
\begin{equation}
P^{\rm (2D)}(e) \approx e/\sqrt{1-e^2},
\label{eq:Pe}
\end{equation}
from which it directly follows that 
\begin{equation}
P^{\rm (2D)}(t_{GW} < \tau) \approx (\tau / t_c)^{1/7},
\label{eq:PtGW}
\end{equation}
where we have used that $P^{\rm (2D)}(e > e_0) = \sqrt{(1-e_0^2)}$. With this expression, we can now estimate the probability that a binary black hole with given orbital parameters
will merge through a $2$-body or a $3$-body merger process.

The probability of a $2$-body merger, $p_2^{\rm (2D)}$, is found by equating $\tau$ in Eq.~\ref{eq:PtGW} with the time until the binary black hole undergoes its next interaction, $t_{\sint}$,
\begin{equation}
p_2^{\rm (2D)} \approx (t_{\sint} / t_c)^{1/7} \approx 0.07 \left[ \frac{t_{\sint}}{{10^{5}}\ {\rm yr}} \right]^{1/7} \left[ \frac{m}{20M_{\odot}} \right]^{3/7} \left[ \frac{a}{1\ {\rm AU}} \right]^{-4/7},
\label{eq:p2}
\end{equation}
where we have normalized to values characteristic of AGN-disk environments\cite{2020ApJ...898...25T}. 
To estimate the $3$-body merger probability, $p_3^{\rm (2D)}$, we describe the often highly chaotic binary-single interaction as a series of $\mathcal{N}$ temporary states, each characterized by a binary black hole with a bound single\cite{2014ApJ...784...71S}. Now, for one of these temporary binary black holes to merge, its inspiral time $t_{GW}$ must be smaller than the characteristic timescale for the system\cite{2017ApJ...846...36S, 2018ApJ...853..140S}, $T_0 \sim \sqrt{a^3/(Gm)}$.
Hence, the probability that a binary black hole undergoes a $3$-body merger, is found by first equating $\tau$ in Eq.~\ref{eq:PtGW} with $T_0$ and then multiplying by $\mathcal{N}$, 
\begin{equation}
p_3^{\rm (2D)} \approx \mathcal{N} \times (T_0 / t_c)^{1/7} \approx 0.15 \left[\frac{m}{20M_{\odot}} \right]^{5/14} \left[\frac{a}{1\ {\rm AU}} \right]^{-5/14}.
\label{eq:p3}
\end{equation}
For the last equality we have used $\mathcal{N} \approx 20$ as found using the simulations introduced in Fig.~\ref{fig:P23_a} (We do not find $\mathcal{N}$ to depend significantly on $\psi$ for
our considered equal-mass setups; however, it does depend on the mass hierarchy and should therefore be treated as a variable in the general case\cite{2018ApJ...853..140S}).
For comparison, in the isotropic 3D case the eccentricity distribution $P(e)$ instead follows\cite{Heggie:1975uy} $P^{\rm (3D)}(e) = 2e$, from which one finds\cite{2018PhRvD..97j3014S},
$p_{2}^{\rm (3D)} \approx (t_{\sint} / t_c)^{2/7}$ and $p_{3}^{\rm (3D)} \approx \mathcal{N} \times (T_0/t_c)^{2/7}$. The relative change in going from 3D to the illustrative 2D case, is therefore
\begin{align}
\frac{p_2^{\rm (2D)}}{p_2^{\rm (3D)}} \approx (t_c/t_{\sint})^{1/7} & \approx 10^1 \times   \left[\frac{m}{20M_{\odot}}\right]^{-6/14}\left[\frac{a}{\rm 1\ {\rm AU}}\right]^{8/14}\left[\frac{t_{\sint}}{10^{5}\ \mathrm{yr}.}\right]^{-1/7}, \\
\frac{p_3^{\rm (2D)}}{p_3^{\rm (3D)}} \approx (t_c/T_0)^{1/7}       & \approx 10^2 \times \left[\frac{m}{20M_{\odot}}\right]^{-5/14}\left[\frac{a}{\rm 1\ {\rm AU}}\right]^{5/14},
\end{align}
which here shows why a disk-like environment can lead to a significant enhancement of both $2$-body- and $3$-body mergers. Comparing the relative probabilities,
\begin{equation}
(p_3^{\rm (2D)}/p_2^{\rm (2D)})/(p_3^{\rm (3D)}/p_2^{\rm (3D)}) \approx \left(t_{\sint}/T_0\right)^{1/7} > 1,
\end{equation}
we further conclude that the relative number of $3$-body mergers forming in the 2D case is always greater than
in the 3D case. This is important in relation to eccentric mergers, as will be described further below.
Lastly, Fig.~\ref{fig:P23_a} illustrates the merger probabilities as a function of $a$, derived using our 2.5-PN $N$-body code\cite{2017ApJ...846...36S},
together with our analytic approximations. 
As seen in this figure, we find excellent agreement in the large $a$ limit as expected, as this is where the probabilities are small
enough to be written as an uncorrelated sum over $\mathcal{N}$ states as we did in Eq.~\ref{eq:p3}.

One immediate consequence of the relative enhancement of $3$-body mergers is a non-negligible population of binary black hole mergers
that appear with a detectable eccentricity, $e_f$, at some gravitational wave frequency, $f_{GW}$. 
As for the merger probabilities above, we start by calculating an upper limit on the number of such eccentric mergers by again considering the illustrative 2D limit.
For these calculations we consider only the contribution from $3$-body mergers as they greatly dominate (see Fig. \ref{fig:fGW_dist}).

For a binary black hole to appear with an eccentricity $e > e_f$ at gravitational wave peak frequency $f_{GW} \approx \pi^{-1}\sqrt{2Gm/r_f^{3}}$, 
where $r_f$ is here the pericenter distance, the
initial binary black hole pericenter distance at assembly (or `capture') has to be smaller than $r_c(e_f)$, where $r_c(e_f)$ at quadrupole-order is given by\cite{2018PhRvD..97j3014S, 2018ApJ...855..124S},
\begin{equation}
r_c(e_f) \approx \left( \frac{2Gm}{f_{GW}^2 {\pi}^2}\right)^{1/3} \frac{1}{2}  \frac{1+e_{f}}{e_{f}^{12/19}} \left[ \frac{425}{304} \left(1 + \frac{121}{304}e_{f}^2 \right)^{-1} \right]^{870/2299}.
\end{equation}
Therefore, the probability that a binary-single interaction results in an eccentric merger during the interaction is, to leading order, the probability that two of the three black holes form
a temporary binary black hole with an initial pericenter $r_0 < r_c(e_f)$.
To calculate this probability, denoted here in short by $p_{ecc}$, we again describe the $3$-body interaction as $\mathcal{N}$ temporary binary-single states, and
by using that $P^{(\rm 2D)}(e>e_0) \approx \sqrt{2(1-e_0)}$ in the high-eccentricity 2D limit, together with the Keplerian relation ${r_0/a} = {1-e_0}$, we find\cite{2018PhRvD..97j3014S},
\begin{equation}
p_{ecc}^{(\rm 2D)} \approx \mathcal{N} \times \sqrt{\frac{2r_{c}(e_f)}{a}}.
\end{equation}
For $f_{GW}=10$\,Hz, relevant for LIGO-Virgo\cite{2019ApJ...871..178G},
\begin{equation}
p_{ecc}^{(\rm 2D)}(e > 0.1: > 10\ {\rm Hz}) \approx 0.15 \left[\frac{m}{20M_{\odot}} \right]^{1/6} \left[\frac{a}{1\ {\rm AU}} \right]^{-1/2},
\end{equation}
where this relation includes all sources that will appear with $e > 0.1$ at $10\ {\rm Hz}$ or above.
Comparing this to the isotropic 3D case\cite{2018PhRvD..97j3014S} for which $p_{ecc}^{(\rm 3D)} \approx \mathcal{N} \times {2r_{c}(e_f)}/{a}$, we find
$p_{ecc}^{(\rm 2D)}/p_{ecc}^{(\rm 3D)} \approx \mathcal{N}/p_{ecc}^{(\rm 2D)}$. Therefore, in the 2D case the probability that a scattering results in an eccentric source is greater by at least 
a factor of $\sim \mathcal{N}$ compared to the 3D case. These analytical considerations are further supported by the scattering results shown in Fig.~\ref{fig:P23_a}.
Finally, note that for converting to the probability for actually observing an eccentric source one has to sum up the individual scattering probabilities -- the ones we have considered so far -- over the
number of scatterings a given binary undergoes before merger takes place. The relative observable number of eccentric mergers is therefore greater
than stated above.

\section{Results from Unequal-mass Scatterings.}

\begin{figure}
\centering
\includegraphics[width=0.95\columnwidth]{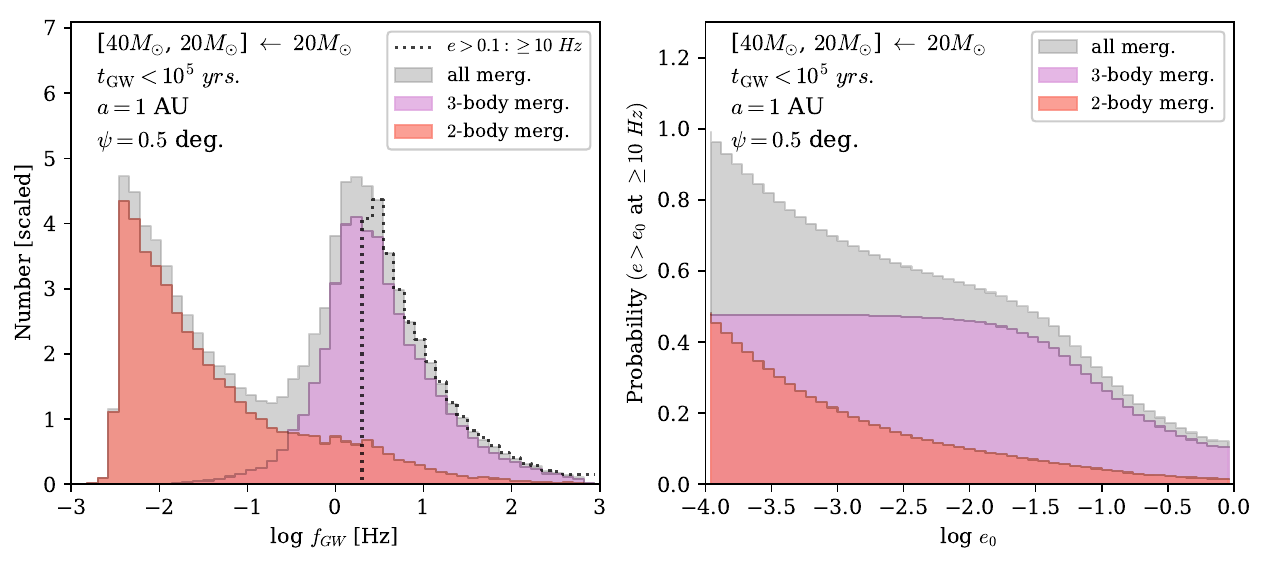}
\caption{{\bf Eccentricity and gravitational wave frequency distributions.} Results from 2.5-PN scatterings,
between an incoming $20M_{\odot}$ black hole and a $[40M_{\odot}, 20M_{\odot}]$ binary black hole with
initial $a = 1\ {\rm AU}$. The initial inclination between the binary and single orbital planes is $\psi = 0.5^\circ$.
The {\it red} and {\it purple} distributions show the outcome from $2$-body mergers and $3$-body mergers with $t_{GW} < t_{\sint} =  10^{5}\ {\rm yr}$, respectively, where {\it grey} is their sum.
{\it Left:} Distribution of binary black hole gravitational wave peak frequency, $f_{GW}$, measured right after the time of assembly.
{\it Right:} Cumulative distribution of binary black hole orbital eccentricities at $\geq 10\ {\rm Hz}$.
For sources formed with $f_{GW} > 10\ {\rm Hz}$, the corresponding eccentricity has been set to $e=1$.
}
\label{fig:fGW_dist}
\end{figure}

To illustrate that eccentric binary black hole mergers are not a unique outcome of the equal-mass interactions mainly presented in this {\it Letter}, we here show results from 2.5-PN-scatterings between a binary black hole [$40M_{\odot}, 20M_{\odot}$] with $a = 1\ {\rm AU}$ and $e=0$, and an incoming single $20M_{\odot}$ black hole on an inclined orbital plane with $\psi = 0.5^\circ$.
Fig.~\ref{fig:fGW_dist} (left panel) shows the gravitational
wave peak frequency distribution at the time of dynamical formation (assembly) for
each binary black hole that eventually merges. Nearly all $3$-body mergers form near or in the
LIGO-Virgo band, whereas the $2$-body mergers peak at lower frequencies in the LISA band.
Fig.~\ref{fig:fGW_dist} (right panel) shows the corresponding cumulative eccentricity distribution at $10$~Hz, where we
have assumed the quadrupole approximation\cite{Peters:1964bc} for propagating the sources from their initial gravitational wave peak frequency.
As seen here, $\sim 40\%$ of all the binaries with merger time $<10^{5}\ \mathrm{yr}$ will appear with an eccentricity $e > 0.1$ at $f_{GW} > 10$~Hz in this unequal-mass example.
This is similar to what is found in the equal-mass case, which illustrates that the unequal-mass case also is expected to produce eccentric mergers at an enhanced rate when the interactions are near co-planar.

\section{Post-Newtonian Scattering Simulations}

We use a 2.5-PN $N$-body code that has been well tested and used in several previous studies\cite{2017ApJ...846...36S}, including
both $3$-body\cite{2018ApJ...853..140S} and $4$-body\cite{2019ApJ...871...91Z} dynamics. Although we had the option of including
all lower-order-PN terms up to 2.5, we only included the 2.5-PN term both to ensure stability\cite{2018PhRvD..98l3005R}, and as this
has been proven sufficient to accurately resolve and track the population of $3$-body mergers that is unique to this channel\cite{2019ApJ...871...91Z}.
We did not include the possible effects from gas-drag in the $3$-body equations-of-motion, or tidal effects from the central super massive black hole. Two-dimensional simulations of binaries in an AGN-disk have been
performed very recently\cite{2021arXiv210109406L}, but an extension to the chaotic three-dimensional few-body problem is highly non-trivial, and will be the topic of future studies.

In all our 2.5-PN simulations we limited the interaction time to $1000$ times the orbital time
of the initial target binary, which resulted in only a few percent of inconclusive scatterings. These mainly originates from scatterings for which the third object is sent out on an almost
unbound orbit with a corresponding orbital time that theoretically approaches infinity. These were therefore excluded from our sample.
In addition, we only considered the strong-scattering-regime, for which the single object on its first passage has an effective peri-center distance with respect to the binary that is
similar to the binary semi-major axis. Therefore, we do not include the regime of weak scatterings\cite{2019PhRvD.100d3010S}, which could take place
during the early stages of the interaction when the single approaches the binary on a near co-rotating orbit as could be the case from the AGN-disk migration model. This naturally requires a more comprehensive
$N$-body modeling of the problem, including effects from the disk and central black hole, and will be addressed in future studies.

For all our presented scatterings based on interactions between a binary and a single on a fixed orbital plane, we assume the singles
have an isotropic distribution of impact parameters at infinity. For the 3D isotropic scattering cases, the distribution is instead taken to be isotropic at
a sphere at infinity. For all interactions, we randomized the internal binary orbital phase-angle, therefore, in this work we do not distinguish 
between counter-rotating and co-rotating interactions. We further assumed the single to nearly free-fall from infinity.

\begin{addendum}
 \item[Data Availability Statement] The data that support the findings of this study are available from the corresponding author upon reasonable request.
\end{addendum}

\end{document}